\documentclass[prbr,aps,twocolumn,groupedaddress]{revtex4}

\usepackage{bm}
\usepackage{graphicx}
\usepackage{color}
\usepackage{ulem}

\begin{document}

\title{Observation of inverted band structure in topological Dirac-semimetal candidate CaAuAs}

\author{Kosuke Nakayama,$^{1,2}$ Zhiwei Wang,$^{3,\ast}$ Daichi Takane,$^1$ Seigo Souma,$^{4,5}$ Yuya Kubota,$^1$ Yuki Nakata,$^1$ Cephise Cacho,$^6$ Timur K. Kim,$^6$ Sandy Adhitia Ekahana,$^7$ Ming Shi,$^7$ Miho Kitamura,$^8$ Koji Horiba,$^8$ Hiroshi Kumigashira,$^{8,9}$ Takashi Takahashi,$^{1,4,5}$ Yoichi Ando,$^3$ and Takafumi Sato$^{1,4,5}$}

\affiliation{$^1$Department of Physics, Tohoku University, Sendai 980-8578, Japan\\
$^2$Precursory Research for Embryonic Science and Technology (PRESTO), Japan Science and Technology Agency (JST), Tokyo, 102-0076, Japan\\
$^3$Institute of Physics II, University of Cologne, K\"{o}ln 50937, Germany\\
$^4$Center for Spintronics Research Network, Tohoku University, Sendai 980-8577, Japan\\
$^5$WPI Research Center, Advanced Institute for Materials Research, Tohoku University, Sendai 980-8577, Japan\\
$^6$Diamond Light Source, Harwell Science and Innovation Campus, Didcot, Oxfordshire OX11 0QX, UK\\
$^7$Swiss Light Source, Paul Scherrer Institut, CH-5232 Villigen, Switzerland\\
$^8$Institute of Materials Structure Science, High Energy Accelerator Research Organization (KEK), Tsukuba, Ibaraki 305-0801, Japan\\
$^9$Institute of Multidisciplinary Research for Advanced Materials (IMRAM), Tohoku University, Sendai 980-8577, Japan
}

\date{\today}

\begin{abstract}
We have performed high-resolution angle-resolved photoemission spectroscopy of ternary pnictide CaAuAs which is predicted to be a three-dimensional topological Dirac semimetal (TDS). By accurately determining the bulk-band structure, we have revealed the coexistence of three-dimensional and quasi-two-dimensional Fermi surfaces with dominant hole carriers. The band structure around the Brillouin-zone center is characterized by an energy overlap between hole and electron pockets, in excellent agreement with first-principles band-structure calculations. This indicates the occurrence of bulk-band inversion, supporting the TDS state in CaAuAs. Because of the high tunability in the chemical composition besides the TDS nature, CaAuAs provides a precious opportunity for investigating the quantum phase transition from TDS to other exotic topological phases.
\end{abstract}

\pacs{71.18.+y, 71.20.-b, 79.60.-i}

\maketitle
Novel quantum states of matter characterized by nontrivial topology of electronic wave functions are an emergent topic in condensed-matter physics \cite{HasanRMP2010,ZhangRMP2011,AndoJPSJ2013}. Among them, topological Dirac semimetals (TDSs) are the first example of a gapless topological phase in the three-dimensional (3D) band structure \cite{MurakamiNJP2007,YoungPRL2012,WangPRB2012,WangPRB2013,LiuScience2014,LiuNM2014,NeupaneNC2014,BorisenkoPRL2015}. In 3D TDSs, an inverted bulk band leads to a point crossing of valence band (VB) and conduction band (CB) to form massless Dirac fermions protected by crystalline symmetries \cite{YoungPRL2012,WangPRB2012,WangPRB2013}. Unlike the two-dimensional (2D) counterpart (graphene), Dirac fermions in 3D TDSs are linearly dispersive along any momentum direction. Such an unusual band structure triggers various intriguing physical properties, as exemplified by high carrier mobility \cite{LiangNM2015}, giant linear magnetoresistance \cite{LiangNM2015,FengPRB2015}, and chiral anomaly \cite{XiongScience2015}. Moreover, TDSs provide a fertile playground for realizing a variety of topological phases such as topological insulators, Weyl semimetals, axion insulators, and topological superconductors, by breaking symmetries, reducing dimensionality, and controlling spin-orbit coupling (SOC) \cite{MurakamiNJP2007,YoungPRL2012,WangPRB2013,KobayashiPRL2015}. Therefore, 3D TDSs are attracting an increasing interest as a suitable platform for exploring exotic topological phenomena and quantum-phase transitions. However, experimental investigations on 3D TDSs are focused mainly on Na$_3$Bi \cite{LiuScience2014} and Cd$_3$As$_2$ \cite{LiuNM2014,NeupaneNC2014,BorisenkoPRL2015} despite many theoretical predictions for other TDS candidates \cite{KariyadoJPSJ2011,GibsonPRB2015,YuPRL2015,CaoPRB2016,LiPRB2017,LePRB2017,ZhangPRB2017,DuQM2017,HunagPRB2017,ZhangJPCL2017,GuoPRB2017,ChangPRL2017,ChenPRM2017}.

Recently, it was proposed that ternary pnictide CaAuAs and its isostructural family, e.g., BaAgAs, are 3D TDSs \cite{SinghPRB2018,MardanyaPRM2019,ChenJPCC2017,TsetserisJPCM2017}. These materials crystalize in the BeZrSi-type structure (space group $P$6$_3/mmc$, No. 194) which has the $C_3$ rotational symmetry with respect to the $c$ axis [see Fig. 1(a) for crystal structure]. First-principles band-structure calculations including SOC have shown that CaAuAs possesses a pair of Dirac cones close to the Fermi level ($E_{\rm F}$) along the $\Gamma$A line of the bulk BZ [see Fig. 1(b)] due to the bulk-band inversion and the protection by the $C_3$ symmetry \cite{SinghPRB2018}. The predicted simple band structure with no $E_{\rm F}$ crossings of other topologically trivial bands makes CaAuAs an ideal system to search for exotic properties associated with bulk Dirac fermions. In addition, the CaAuAs family has a great potential as a parent compound of different types of topological states. For example, the reduction of SOC, e.g., by replacing As with P, would cause a topological phase transition into a nodal-line semimetal (NLS) characterized by a continuous band crossing along 3D curves with a starfruit-like shape \cite{SinghPRB2018}, distinct from a one-dimensional shape of the nodal lines in typical NLSs \cite{SchoopNC2016,TakanePRB2016,TakaneQM2018}. It was also predicted that, by breaking crystalline symmetries \cite{SinghPRB2018,MardanyaPRM2019}, this system is transformed into Weyl semimetal or three-fold-fermion state. Moreover, superconductivity can be introduced by the replacement of As with Bi \cite{XuJGG2020}. Despite such interesting proposals and known physical properties, there exist no experimental report on the band structure of CaAuAs family. It is thus desirable to clarify its fundamental band structure.

In this Rapid Communication, we report angle-resolved photoemission spectroscopy (ARPES) of CaAuAs single crystal. By utilizing energy-tunable photons from synchrotron radiation, we experimentally established the electronic structure over the 3D bulk BZ, and found evidence for the bulk-band inversion which is a prerequisite for realizing the 3D TDS state. We compare the obtained results with the first-principles band-structure calculations and discuss implications in relation to the topological properties.

\begin{figure}
\includegraphics[width=3.4in]{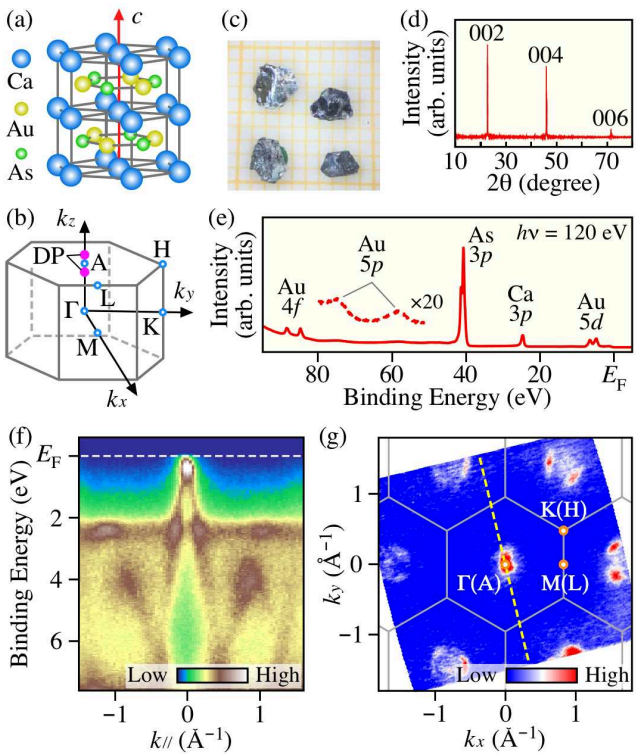}
\vspace{0cm}
\caption{(Color online) (a) and (b) Crystal structure and bulk hexagonal BZ of CaAuAs, respectively. Magenta dots in (b) denote the \textbf{k} points where the Dirac point (DP) exists. (c) and (d) Typical photograph and X-ray diffraction patterns of CaAuAs, respectively. (e) EDC in a wide energy range measured at $h\nu$ = 120 eV. (f) ARPES intensity plot in the VB region measured along a yellow dashed line in (g) at $T$ = 30 K at $h\nu$ = 120 eV. It is noted that uncertainty in the azimuthal angle of the \textbf{k} cut is fairly small (less than 2$^{\circ}$) because the band-structure mapping over a wide momentum space enables us to precisely determine the azimuthal angle based on the periodicity and symmetry of the band structure. (g) ARPES intensity map at $E_{\rm F}$ as a function of 2D wave vector ($k_x$ and $k_y$) obtained by integrating the spectral intensity within $\pm$20 meV of $E_{\rm F}$.}
\end{figure}

CaAuAs single crystals [typical size of 2 $\times$ 3 $\times$ 0.5 mm$^3$; Fig. 1(c)] were synthesized by a self-flux method. High purity Ca shot, Au powder, and As lump were loaded in an Al$_2$O$_3$ crucible with the mole ratio of Ca:Au:As = 1:3:3. Then the crucible was sealed in a quartz tube in a high vacuum of 4 $\times$ 10$^{-6}$ Torr. The tube was heated to 1323 K in 20 h, kept at that temperature for another 20 h, and then slowly cooled down to 873 K at a rate of 5 K/h. Finally, the tube was cooled down to room temperature. X-ray diffraction measurements [Fig. 1(d)] confirm that the grown crystal is of high structural quality and in the target phase. ARPES measurements were performed with Scienta-Omicron electron analyzers at the beamline I05 in DIAMOND Light Source (DLS) \cite{HoeschRSI2017}, BL-2A in Photon Factory (PF), KEK, and SIS X09LA in Swiss Light Source (SLS), PSI. We used energy-tunable photons of 95-120 eV in DLS, 60-185 eV in PF, and 24-90 eV in SLS. The energy and angular resolutions were set to be 15-30 meV and 0.2-0.3$^{\circ}$, respectively. Crystals were cleaved $in$ $situ$ along the (001) crystal plane in an ultrahigh vacuum of better than 1 $\times$ 10$^{-10}$ Torr (see Supplemental Material for the characterization of cleaved surface \cite{SM}). The sample was kept at $T$ = 30-40 K during the ARPES measurement. Figure 1(e) displays the energy distribution curve (EDC) in a wide energy range measured at photon energy ($h\nu$) of 120 eV. One can recognize several core-level peaks originating from the Ca (3$p$), Au (4$f$, 5$p$, 5$d$), and As (3$p$) orbitals. The absence of core levels from other atomic species confirms the clean sample surface.

First, we show the overall VB structure. Figure 1(f) displays the VB structure along a \textbf{k} cut crossing the $\Gamma$(A) point measured at $T$ = 30 K with 120-eV photons. There are several energy bands such as a highly dispersive holelike band touching $E_{\rm F}$ at the wave vector $k_{\parallel}$ $\sim$ 0, another holelike band topped at a binding energy ($E_{\rm B}$) of $\sim$4 eV at the BZ boundary ($k_{\parallel}$ $\sim$ 0.9 \AA$^{-1}$), and a relatively flat band at $E_{\rm B}$ $\sim$ 2.5 eV. These energy bands have a dominant contribution from the As 4$p$ orbitals \cite{SinghPRB2018}. Figure 1(g) shows the ARPES-intensity mapping at $E_{\rm F}$ as a function of 2D wave vector. One can immediately recognize that the observed intensity pattern follows well the periodicity of the hexagonal BZ. As expected from the $E_{\rm F}$ touching of a band in Fig. 1(f), there is a small Fermi surface (FS) centered at the $\Gamma$(A) point in the first BZ, without indication of any other FSs away from the $\Gamma$(A) point. One can also identify a FS at the $\Gamma$(A) point in the second BZs, although the size is larger than that in the first BZ due to the 3D character as we demonstrate later.

\begin{figure}
\includegraphics[width=3.4in]{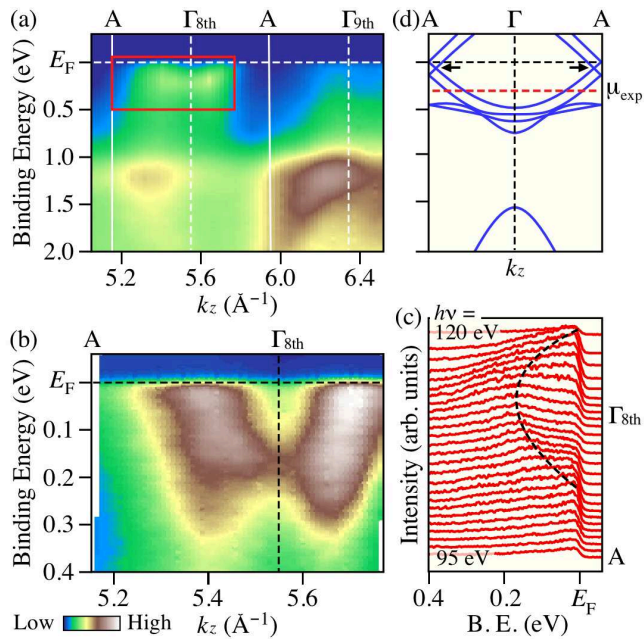}
\vspace{0cm}
\caption{(Color online) (a) Normal-emission ARPES intensity plotted as a function of $k_z$ for CaAuAs. The inner potential $V_0$ was estimated to be 11.5 eV from the periodicity of the band dispersion. (b) and (c) High-resolution near-$E_{\rm F}$ ARPES intensity and corresponding EDCs, respectively, obtained in the ($E$, \textbf{k}) region enclosed by a red rectangle in (a). Black dashed curve in (c) is a guide for the eyes to trace the electronlike band dispersion. (d) Calculated bulk-band structure along the $\Gamma$A high-symmetry line extracted from the first-principles band-structure calculation with SOC in ref. 29. Black arrows indicate the location of Dirac points.}
\end{figure}

Next we have performed normal-emission ARPES measurements at various $h\nu$'s and determined the band structure along the wave vector perpendicular to the sample surface ($k_z$). In Fig. 2(a), one can see an intensity variation as a function of $k_z$. For example, strong intensity appears near $E_{\rm F}$ around each $\Gamma$ point ($k_z$ $\sim$ 5.6 and 6.4 \AA$^{-1}$ corresponding to the $\Gamma$ point of the 8th and 9th BZ, respectively). This strong intensity originates from an electron pocket with the bottom of the dispersion at $E_{\rm B}$ $\sim$ 200 meV, as confirmed by high-resolution measurements in Figs. 2(b) and 2(c). One can also recognize a holelike dispersion topped at the $\Gamma$ point at higher $E_{\rm B}$ ($\sim$1.3 eV) [see around $\Gamma_{9th}$ in Fig. 2(a)]. The clear dispersion along the $k_z$ direction demonstrates the bulk origin of the observed bands. According to first-principles band-structure calculations including SOC [Fig. 2(d)] \cite{SinghPRB2018}, the topmost VB and a higher-energy band at $E_{\rm B}$ $>$ $\sim$ 1 eV have the bottom and top of the dispersion at $\Gamma$, respectively, qualitatively consistent with the present observation. However, there is a clear difference between the experiment and calculation; the topmost VB touches $E_{\rm F}$ in the vicinity of the A point in the calculation [Fig. 2(d)], whereas it crosses $E_{\rm F}$ midway between the $\Gamma$ and A point in the experiment [Figs. 2(b) and 2(c)]. This quantitative difference suggests that the measured CaAuAs crystal is hole-doped. By comparing the energy position of the electron-band bottom between the experiment and calculation, the doping-induced shift of the chemical potential is estimated to be $\sim$300 meV, as indicated by a red dashed line in Fig. 2(d). The upward shift of the holelike band at $E_{\rm B}$ $\sim$ 1 eV in the experiment [compare Figs. 2(a) and 2(d)] is also explained in terms of the hole doping and the resultant chemical-potential shift. It is noted that, while there are four bands within 1 eV of $E_{\rm F}$ in the calculation [Fig. 2(d)], only one band is resolved in Figs. 2(a)-2(c). This is possibly due to the $k_z$-broadening effect and/or the matrix-element effect of photoelectron intensity.

\begin{figure*}
\includegraphics[width=6.8in]{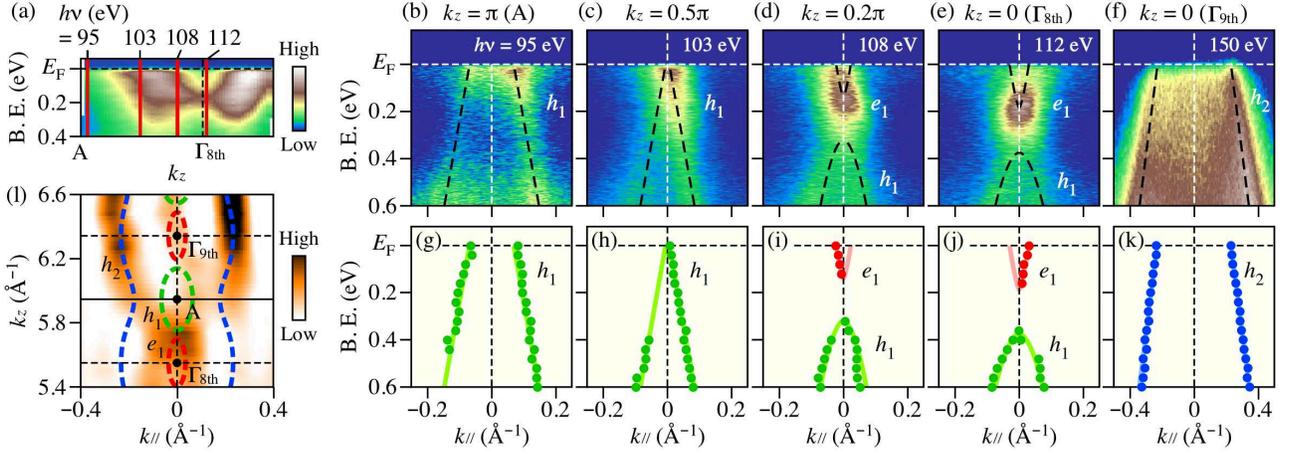}
\vspace{0cm}
\caption{(Color online) (a) Same as Fig. 2(c). (b)-(f) $h\nu$ dependence of near-$E_{\rm F}$ APRES intensity measured along a yellow dashed line in Fig. 1(f), and (g)-(k) corresponding band dispersions determined by tracing the peak position of MDCs. (l) ARPES-intensity map at $E_{\rm F}$ as a function of $k_z$ and $k_x$. The intensity was obtained by integrating the second-derivative MDCs within $\pm$20 meV of $E_{\rm F}$. Dashed curves show the experimental FSs determined by smoothly connecting the Fermi wave vectors for the $h_1$, $h_2$, and $e_1$ bands.}
\end{figure*}

To clarify the fermiology of CaAuAs in 3D BZ in more detail, we plot the ARPES intensity [Figs. 3(b)-3(f)] and the corresponding band dispersion [Figs. 3(g)-3(k)] at several different $k_z$ values. As seen in Figs. 3(b) and 3(g), a holelike band (hereafter we call $h_1$) crosses $E_{\rm F}$ to form a FS centered at the A point at $k_z$ = $\pi$. This band moves toward higher $E_{\rm B}$ on going away from $k_z$ = $\pi$ and sinks below $E_{\rm F}$ near $k_z$ $\sim$ 0.5$\pi$ [Figs. 3(c) and 3(h)]. With further approaching $k_z$ = 0, the $h_1$ band further shifts downward and an electronlike band ($e_1$) appears [Figs. 3(d) and 3(i)]. Finally, at $k_z$ = 0, the bottom of $e_1$ band and the top of $h_1$ band reaches $E_{\rm B}$ $\sim$ 0.2 and 0.4 eV, respectively [Figs. 3(e) and 3(j)]. The electronlike dispersion in Figs. 2(a)-2(c) mainly reflects the $e_1$ band, as inferred from the much stronger intensity than the $h_1$ band in Fig. 3. In addition to the $h_1$ and $e_1$ bands, our $h\nu$-dependent study over a wide $k_z$ region uncovered another holelike band ($h_2$) in the next (9th) BZ [Figs. 3(f), 3(k), and 3(l)]. As visible in Figs. 3(f) and 3(k), the $h_2$ band has a relatively large Fermi wave vector at $k_z$ = 0 compared with that of the $h_1$ or $e_1$ band, and is slightly dispersive along the $k_z$ direction, forming a cylindrical FS with a finite wiggling, as visualized in the $k_z$-$k_x$ FS mapping in Fig. 3(l) (blue dashed curves).  It is remarked here that, although intensity of the $h_2$ band is strongly suppressed in the 8th BZ [Figs. 3(b)-3(e) and 3(l)] due to the matrix-element effect, its existence as well as its quasi-2D character is supported by the ARPES measurements at different conditions [the measurement also revealed a signature of additional holelike band ($h_3$)]; for details, see Supplemental Material \cite{SM}.

\begin{figure}
\includegraphics[width=3.4in]{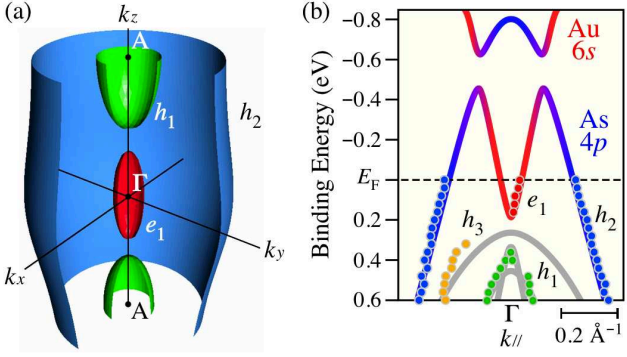}
\vspace{0cm}
\caption{(Color online) (a) Schematic 3D view of the FS for CaAuAs deduced from the present ARPES experiment. (b) Experimental band dispersion around the $\Gamma$ point obtained by overlapping the results in Figs. 3(j), 3(k), and supplemental Fig. S2(i) (filled circles) \cite{SM}, compared with the calculated band dispersion with SOC along the $\Gamma$M line (solid curves) \cite{SinghPRB2018}.}
\end{figure}

As schematically drawn in Fig. 4(a), the present ARPES measurements revealed that the FS of CaAuAs consists of 3D electronlike ($e_1$) and holelike ($h_1$) pockets centered at the $\Gamma$ and A points, respectively, as well as a quasi-2D holelike ($h_2$) cylinder along the $\Gamma$A line [here we assumed an isotropic radius in the $k_x$-$k_y$ plane based on a circular shape of the FS in the second BZs in Fig. 1(g)]. The formation of a quasi-2D cylindrical FS is probably associated with the hole doping to the crystal. In fact, the carrier concentration estimated from the FS volume (the size of $e_1$, $h_1$, and $h_2$ pockets is 0.05, 0.2, and 7.0$\%$ of bulk BZ, respectively) is $\sim$8 $\times$ 10$^{20}$ cm$^{-3}$ holes, which roughly agrees with the Hall-coefficient measurements (3 $\times$ 10$^{20}$ cm$^{-3}$ holes). After taking the hole doping into account, we found a good agreement between the experimental and calculated bulk bands. In particular, the band dispersion at $k_z$ = 0 obtained by combining the results in Figs. 3(e), 3(f), and supplemental Fig. S2(i) \cite{SM} coincides nearly perfectly with the calculation in which the chemical potential is shifted downward by 300 meV \cite{SinghPRB2018}. The present result demonstrates that the bulk-band structure of CaAuAs is well captured by the band-structure calculations.

Now we discuss the implications of present results in relation to the topological properties of CaAuAs. An important finding is the nontrivial band topology. In CaAuAs or related compounds with the same crystalline symmetry, the VB with the pnictogen $p$-orbital character exhibits the band maximum at $\Gamma$; e.g., in the case of trivial semiconductor BaAgP, all the near-$E_{\rm F}$ VB show a holelike dispersion centered at $\Gamma$ \cite{SinghPRB2018,MardanyaPRM2019}. Importantly, the electronlike dispersion shows up below $E_{\rm F}$ only when the CB bottom (Au $s$ orbital) crosses the VB top to form a nontrivial inverted band structure \cite{SinghPRB2018,MardanyaPRM2019}. Therefore, the present observation of electronlike pocket $e_1$ at $\Gamma$ provides compelling evidence for the occurrence of bulk-band inversion in CaAuAs, as illustrated in Fig. 4(b). When the band inversion takes place in the absence of SOC, the hybridization between VB and CB is prohibited on the three equivalent $\Gamma$MLA mirror planes owing to the different mirror eigenvalues, leading to crossed nodal rings with a 3D starfruit-like shape \cite{SinghPRB2018}. After the SOC is turned on (this is expected for actual material), the most part of nodal ring is gapped out [see a clear hybridization-gap opening at $E_{\rm B}$ $\sim$ -0.5 eV in Fig. 4(b)] and the band crossing is protected only on the $\Gamma$A line by the $C_3$ rotational symmetry, leaving a pair of Dirac-point nodes near the A point [see black arrows in Fig. 2(d)] \cite{SinghPRB2018}. The inverted band structure clarified in this study thus supports the TDS phase in CaAuAs. A next important challenge is to directly observe the crossing of VB and CB at the Dirac point. Since CaAuAs is hole-doped, the Dirac points are likely located at $\sim$250 meV above $E_{\rm F}$ [Fig. 2(d)], so that the sample with a suppressed doping level is highly desired to firmly establish the TDS nature and to explore novel quantum phenomena associated with the Dirac fermions.

Finally, we compare the characteristics between CaAuAs and well-known TDSs, Na$_3$Bi and Cd$_3$As$_2$. In the electronic-structure point of view, these compounds share a common feature that the Dirac points are protected by the rotational symmetry. However, there exist several important differences among them. First, the Dirac point appears on the [001] ($k_z$) axis in CaAuAs and Na$_3$Bi \cite{LiuScience2014} but it is on the [111] axis in Cd$_3$As$_2$ \cite{LiuNM2014,NeupaneNC2014,BorisenkoPRL2015}. Second, the Dirac point is located near the A point in CaAuAs whereas it is around the $\Gamma$ point in Na$_3$Bi \cite{LiuScience2014}. Third, CaAuAs is rather chemically stable in air, as opposed to highly reactive Na$_3$Bi. Last, CaAuAs has a better tunability of constituent elements by chemical substitution, as recognized from a wide variety of isostructural sister compounds \cite{XuJGG2020}. Such a wide tunability in CaAuAs is useful for controlling the strength of SOC and triggering distinct quantum phases such as superconducting and NLS states. Therefore, the CaAuAs family would provide a suitable platform for investigating the interplay between Dirac fermions and exotic topological phases.

In conclusion, we reported the ARPES results on CaAuAs to clarify the band structure relevant to the nontrivial band topology. We have successfully determined the FS topology and the bulk-band dispersions in the 3D momentum space, and found an excellent agreement between the experimental and calculated bulk-band structures. Most notably, the observed electron pocket at the $\Gamma$ point is a hallmark of the bulk-band inversion, supporting the emergence of a 3D TDS phase in CaAuAs.

\begin{acknowledgments}
We thank K. Sugawara, K. Hori, and T. Kato for their assistance in the ARPES measurements. We also thank Diamond Light Source for access to beam line I05 (Proposal number: SI23799), KEK-PF for beam line BL2A (Proposal number: 2018S2-001 and 2018G-653), and Swiss Light Source for beam line SIS X09LA (Proposal number: 20181888). This work is supported by JST-PRESTO (No. JPMJPR18L7), JST-CREST (No. JPMJCR18T1), Grant-in-Aid for Scientific Research on Innovative Areas ``Topological Materials Science" (JSPS KAKENHI Grant Numbers JP15H05853 and JP15K21717), Grant-in-Aid for Scientific Research (JSPS KAKENHI Grant Numbers JP17H01139, JP18H04472, JP18H01160, and JP17H04847), Grant-in-Aid for JSPS Research Fellow (No. JP18J20058), and the Deutsche Forschungsgemeinschaft (DFG, German Research Foundation) - Project No. 277146847 - CRC 1238 (Subproject A04).
\end{acknowledgments}

\newpage
\bibliographystyle{prsty}

\end{document}